\documentstyle[aps,multicol]{revtex}
\begin{document}
\title{Kinetic Theory of Traffic Flows}
\author{E.~Ben-Naim$^1$ and P.~L.~Krapivsky$^2$}
\address{$^1$Theoretical Division and Center for Nonlinear Studies,
Los Alamos National Laboratory, Los Alamos, New Mexico 87545}
\address{$^2$Center for Polymer Studies and Department of Physics,
Boston University, Boston, Massachusetts 02215} 
\maketitle
\begin{abstract}

We describe traffic flows in one lane roadways using kinetic theory,
with special emphasis on the role of quenched randomness in the
velocity distributions.  When passing is forbidden, growing clusters
are formed behind slow cars and the cluster velocity distribution is
governed by an exact Boltzmann equation which is linear and has an
infinite memory.  The distributions of the cluster size and the
cluster velocity exhibit scaling behaviors, with exponents dominated
solely by extremal characteristics of the intrinsic velocity
distribution. When passing is allowed, the system approaches a steady
state, whose nature is determined by a single dimensionless number,
the ratio of the passing time to the collision time, the two time
scales in the problem. The flow exhibits two regimes, a laminar flow
regime, and a congested regime where large slow clusters dominate the
flow.  A phase transition separates these two regimes when only the
next-to-leading car can pass.
\end{abstract}
\begin{multicols}{2}
\section{Introduction}

Traffic flows are strongly interacting many-body systems.  Therefore,
theoretical techniques such as kinetic theory and hydrodynamics are
useful in describing the rich phenomenology of traffic flows which
includes shock waves, phase transitions, clustering, metastability,
hysteresis, etc.  Traffic is typically modeled within macroscopic
descriptions such as hydrodynamics and kinetic theory
\cite{hydr1,hydr2,Prigogine,Pav,kerner} or microscopic approaches,
e.g., cellular automata
\cite{Nagel,Biham,Nagatani,Schadschneider,Nagel1,Nagel2,Brankov} and
car-following models \cite{pipes,jap1,jap2,jap3,Nagatani1}. The large
body of recent work on the physics of traffic flows is surveyed in
Refs.~\cite{wolf2,css,dh}.

In this review, we describe how quenched randomness in the car
velocities leads to formation of clusters in one lane roadways.  We
assume ballistic motion with infinite memory, namely that each car has
a preferred ``intrinsic'' velocity by which it drives in the absence
of other cars \cite{newell,bkr,bk1,bk2,bk3,ik1}. While the emerging
behavior is quite similar to that found in stochastic particle hopping
processes with quenched disorder
\cite{Nagatani2,Benjamini,krug,Evans,Ktitarev,Krug,Krug2}, these
ballistic motion models often admit deeper analytical treatment.  Our
starting point is an idealized no passing flow, where an exact
analytical solution is possible, and an exact kinetic theory can be
constructed \cite{bkr}.  We then treat more realistic generalizations
where passing is allowed using approximate kinetic theories
\cite{bk1,bk2,bk3,ik1}.  Our goal is to provide a concise summary
where key features are emphasized and outstanding open issues are
highlighted.

\section{No passing zones}

Our basic traffic model mimics cluster formation (often also called
platoon formation) in one-lane roadways where passing is forbidden
\cite{newell,bkr}.  In this model, each car moves ballistically at its
initial velocity until it overtakes the preceding car or cluster.
After this encounter, the incident car assumes the velocity of the
cluster which it has just joined. Cars are taken to be size-less, and
collisions to be instantaneous.  We primarily consider spatially
homogeneous situations where the positions and the velocities of the
cars are initially uncorrelated.  Specifically, cars are distributed
randomly in space with a concentration $c_0$, and their velocities are
drawn from the initial velocity distribution $P_0(v)$.  Remarkably,
analytic expressions can be obtained for the velocity distribution and
the joint size-velocity distribution of clusters.  Furthermore, it is
also possible to describe analytically spatial inhomogeneities and
even input of cars into the system.

\subsection{The velocity distribution}

We first consider the cluster velocity distribution. In this description,
only the lead car is relevant and trailing cars in a cluster can be ignored.
Let $P(v,t)$ be the distribution of clusters with velocity $v$ at time $t$.
Initially, all cars are lead cars and the cluster (or lead-car) velocity
distribution equals $P(v,t)=P_0(v)S(v,t)$, a product of the initial velocity
distribution and the survival probability $S(v,t)$. The survival probability
is the probability that a car with initial velocity $v$ avoids ``collisions''
with slower cars up to time $t$, and hence, is still moving at the same
velocity. Consider a car of initial velocity $v$.  To ensure that it would
not overtake slower cars of velocity $v'<v$, an interval of size $(v-v')t$
ahead of it must not contain $v'$-cars initially. For the velocity
distribution, $P_0(v)$, and a Poissonian initial spatial distribution, the
probability for such an event is $\exp\left[-t\,(v-v')P_0(v')\right]$. For a
car to survive to time $t$, this exclusion probability should be taken into
account for every $v'<v$, and taking the product over all possible slower
cars yields the survival probability, $S(v,t)=\exp[-t\int_0^v
dv'(v-v')P_0(v')]$. Consequently, the cluster velocity distribution is found
for arbitrary initial conditions
\begin{equation}
\label{pvt}
P(v,t)=P_0(v)\exp\left[-t\int_0^v dv'(v-v')P_0(v')\right].
\end{equation}
The process is deterministic (the initial condition is the only
source of randomness) and given the initial positions and velocities
of the cars, the state of the system at any later time follows. This
is reflected nicely in Eq.~(\ref{pvt}).

The exact solution (\ref{pvt}) satisfies the following Boltzmann equation,
\begin{equation}
\label{exact}
{\partial P(v,t)\over \partial t}=-P(v,t)\int_0^v dv'(v-v')P(v',0).
\end{equation}
Interestingly, this rate equation is {\em linear} in the velocity
distribution $P(v,t)$. The collision rate is proportional to the
relative velocity, $v-v'$, and the {\em initial} velocity distribution
of the slower cars, a signature of the infinite memory in the system.
This is a unique case where the hierarchy of evolution equations
corresponding to the velocity distributions terminates at the first
order.

Whereas the steady state of one-lane traffic with no passing is trivial,
viz. all cars will eventually join a cluster led by the slowest car in the
system, the time dependent behavior is interesting. We concentrate on the
long time behavior, which is largely independent of the initial distribution
of fast cars, as follows from Eq.~(\ref{pvt}). For discrete velocity
distributions, the time dependence is exponential, and we focus on continuous
distributions. In this case, we find directly from Eq.~(\ref{pvt}) that both
the cluster concentration, \hbox{$c(t)=\int\, dv P(v,t)$}, and the average
velocity, \hbox{$\langle v(t)\rangle=c^{-1}\int\, dv\, v\, P(v,t)$}, decay
algebraically with time
\begin{eqnarray}
\label{cvscl}
c(t)&\sim& t^{-\alpha}\qquad\qquad \alpha={\mu+1\over \mu+2},\nonumber\\
\langle v(t)\rangle&\sim& t^{-\beta}\qquad\qquad\beta={1\over \mu+2}.
\end{eqnarray}
The scaling exponents $\alpha$ and $\beta$ depend only on the small-$v$
extremal statistics \cite{nextreme} of the initial velocity distribution,  
\begin{equation}
\label{piv}
 P_0(v)\simeq av^{\mu}\qquad v\to 0,
\end{equation}
via the cutoff exponent $\mu>-1$. The two scaling exponents are related by
$\alpha+\beta=1$ as dictated by an elementary mean free path argument:
$cvt\sim 1$.

Since the number of cars is conserved, the average cluster size is
inversely proportional to the concentration, $\langle m\rangle\sim
1/c$, and the size growth law is $\langle m\rangle\sim t^{\alpha}$.
In the limit $\mu\to \infty$, the size grows linearly with time.  In
contrast, when $\mu\to -1$, the size remains roughly constant, since
the velocity distribution becomes effectively unimodal and collisions
become exceedingly rare. This qualitative dependence on the form of
the initial velocity distribution is reminiscent of ballistic
annihilation processes, where ballistically moving particles
annihilate upon collision \cite{ef,brl,drfp,ksi}. The above clustering
process can be viewed as a ballistic aggregation process that
possesses a single mass conservation law.  The sensitive dependence on
the initial conditions is in contrast with momentum conserving
ballistic agglomeration processes (that mimics large scale formation
of matter in the universe) where a universal scaling
asymptotic behavior emerges \cite{sz,cpy,f}.

The average velocity is the only relevant velocity scale in the
problem and asymptotically the velocity distribution follows the
scaling form
\begin{equation}
P(v,t)\simeq t^{\beta-\alpha}\Phi(vt^{-\beta}).
\end{equation}
{}From Eqs.~(\ref{pvt}) and (\ref{piv}), the scaled distribution is
$\Phi(z)=a z^{\mu} \exp(-bz^{\mu+2})$, with \hbox{$b=a/[(\mu+1)(\mu+2)]$}.
Therefore, the small-$v$ asymptotics of the initial velocity distributions
governs not only the scaling exponents but also the entire shape (including
the large velocity tail) of the scaling function $\Phi(z)$. This scaling
behavior indicates that at time $t$, most cars moving initially with
velocities larger than the typical velocity scale, $\langle v(t)\rangle \sim
t^{-\beta}$, have already joined clusters led by slower cars, while cars
slower than this velocity scale are still driving with their initial
velocity. 

\subsection{The size-velocity distribution}

Given the nature of the model, a car is only affected by the initial
configuration of cars {\it ahead} of it. This key feature enables
solution of the cluster velocity distribution and it allows treatment
of a more detailed quantity, the joint size-velocity
distribution. To obtain $P_m(v,t)$, the density of clusters of size
$m$ and velocity $v$, it is useful to introduce the
cumulative distribution, $Q_m(v,t)$, the distribution of clusters of
velocity $v$ containing at least $m$ cars. Knowledge of this
cumulative distribution yields the joint size-velocity distribution
via differencing, $P_m(v,t)=Q_m(v,t)-Q_{m+1}(v,t)$.

Consider the first nontrivial quantity, $Q_2(v,t)$, the probability
distribution of clusters of velocity $v$ with at least two cars at
time $t$.  This quantity is equal to the product of the probability
that lead car has survived up to time $t$, $P(v,t)$, and the
probability that the car trailing it actually experiences a collision
prior to time $t$.  Let $x_1$ and $v_1$ be the initial position and
the initial velocity of the trailing car, respectively.  For such a
collision to occur, the trailing car must be faster than the lead car,
$v_1>v$, and the interval separating the two cars must be initially
free of other cars.  The probability for this composite event is the
product of the probabilities of each individual event. Given a random
(Poisson) spatial distribution, the probability an interval is empty
is exponential in its length, and the collision probability is
\begin{equation}
\label{q2vt}
Q_2(v,t)=P(v,t)\int_{v}^{\infty} dv_1 P_0(v_1)
\int_{x_1<(v_1-v)t}dx_1\exp(-x_1).
\end{equation}
The fact that the trailing car cannot be slowed down by any other car
before colliding with the lead car is crucial in obtaining this
solution. This solution can be generalized to arbitrary cluster sizes.
Following the two-car cluster case, one simply integrates over all the
initial positions and velocities of the consecutive cars to eventually
collide with the lead car in the cluster to give
\begin{equation}
\label{qmvt}
Q_m(v,t)=P(v,t)\prod_{i=1}^{m-1}\!\!\!\! \int_{v}^{\infty}dv_i\,P_0(v_i)\!\!\!\!\!
\int\limits_{x_1+\cdots+x_i<(v_i-v)t}\!\!\!\!\!dx_i\exp(-x_i).
\end{equation}
The integration limits reflect the fact that all the colliding cars
must be faster than the lead car, and the restriction on the
integration limits ensures that cars are sufficiently close to the
lead car so that collisions indeed occur.

Given the cumulative car distribution, the joint size-velocity distribution
can be formally obtained.  Since $\langle m\rangle\sim t^\alpha$ and $\langle
v\rangle\sim t^{-\beta}$, we anticipate the following scaling behavior:
$P_m(v,t)\simeq t^{\beta-2\alpha} \Psi(mt^{-\alpha},vt^{\beta})$.  This
indeed holds and from Eq.~(\ref{qmvt}) one obtains the scaled joint
distribution
\begin{equation}
\label{phixz}
\Psi(x,z)= c z^\mu (x+z)^{\mu+1}\exp\left[{-b(x+z)^{\mu+2}}\right],
\end{equation}
with $c=a^2/(\mu+1)$.  Again only two parameters, $a$ and $\mu$
characterizing the small velocity characteristics of the initial
conditions, are needed to fully describe the asymptotic state of the
system.  The joint distribution (\ref{phixz}) provides a comprehensive
description of the traffic clustering process. It may be considered as
the counterpart of the well-known result for diffusion-controlled
aggregation in one-dimension \cite{s}.

Integration of the scaling function with respect to $x$ reproduces the scaled
velocity distribution $\Phi(z)$. The complementary scaled size distribution
cannot be found in a closed elementary form, except for the special case of
asymptotically flat distributions ($\mu=0$) where both of the single variable
scaling functions are purely Gaussian.

\subsection{Generalizations}

A natural generalization is to spatially heterogeneous initial
velocity distributions, $P_0(x,v)$. The time and space dependent
cluster velocity distribution, $P(x,v,t)$, follows from a
straightforward generalization of the basic derivation in the
homogeneous case
\begin{equation} 
P(x,v,t)=P_0(x-vt,v) \exp
\left[{-\int_0^vdv'\int_{x-vt}^{x-v't} dx'P_0(x',v')}\right].
\end{equation} 
For instance, consider the special case where cars are uniformly distributed
in the region $x\leq 0$ while the region ahead is empty.  Here, one finds a
governing length scale $x\sim v t\sim t^{\alpha}$ with the same exponent
$\alpha$ as in Eq.~(\ref{cvscl}). This length scale characterizes a
propagating front of clusters, and the space dependent concentration $c(x,t)$
becomes a function of the scaling variable $X=xt^{-\alpha}$, namely
$c(x,t)=t^{-\alpha} {\cal C}(X)$.  Far from the origin, the scaled density
decays as ${\cal C}(X)\sim
X^{-1}$ implying $c(x,t)\sim x^{-1}$ for $x\gg t^{\alpha}$. Consequently, the
total number of clusters in the originally empty region, $N(t)=\int_0^\infty
dx\, c(x,t)$, grows logarithmically slow with time
\begin{equation} 
N(t)\sim \ln t.
\end{equation} 
This growth law is universal as the dependence on the details of the
initial velocity distribution is secondary, entering only via the
prefactor.

In summary, a scaling asymptotic behavior characterizes the kinetics
of clustering in no-passing zones of one lane roadways. The
corresponding scaling exponents and scaling functions are
characterized by the small-velocity statistics of the initial velocity
distributions. Remarkably, it is possible to derive the exact
Boltzmann equation in this case.

\section{Passing zones}

We now describe the complementary case of passing zones where fast
cars can pass slow cars. The model we consider is a straightforward
generalization of the no-passing case. The initial conditions are
identical: cars are distributed randomly in space with concentration
$c_0$ and their velocity is drawn from the intrinsic velocity
distribution $P_0(v)$. The characteristic velocity scale is taken to
be $v_0$. In the absence of other cars, cars drive ballistically with
their intrinsic velocity. In the presence of other cars, two competing
mechanism may cause a change in the car velocity. Collisions lead to
slowing down: when a cluster overtakes a slower cluster, a larger
cluster moving with the smaller of the two velocities forms. Passing
leads to a velocity increase: every car inside a cluster may
spontaneously pass the lead car and resume driving with its intrinsic
velocity. The corresponding passing rate equals a constant,
$t_0^{-1}$.  This is a significant simplification; in realistic
situations only the first few trailing cars may be able to pass.

It proves convenient to introduce dimensionless velocity, space, and time
variables: $v/v_0\to v$, $xc_0\to x$, $c_0v_0t \to t$. This rescales the
passing rate, $t_0^{-1}\to R^{-1}$, where
\begin{equation}
\label{R}
R={t_{\rm pas}\over t_{\rm col}}=c_0v_0t_0
\end{equation}
is the ratio of the passing time $t_{\rm pas}=t_0$ to the collision time
$t_{\rm col}=(c_0v_0)^{-1}$.  We term this fundamental dimensionless quantity
the ``collision number'' and denote it $R$ as it is reminiscent of the
Reynolds number -- the small $R$ limit is straightforward and the large $R$
limit is characterized by boundary layers.

The starting point for kinetic theory is again the cluster velocity
distribution $P(v,t)$.  The approximate Boltzmann equation reads
\begin{eqnarray}
\label{pvt-p}
{\partial P(v,t)\over \partial t}&=&R^{-1}\left[P_0(v)-P(v,t)\right]\\
&-&P(v,t)\int_0^v dv' (v-v')P(v',t).\nonumber
\end{eqnarray}
This evolution equation assumes molecular chaos, namely that the stochastic
passing events effectively mix the velocities, and therefore, spatial
correlations can be neglected.  This is clearly an approximation, as the
collision integral does not coincide with the exact collision integral
(\ref{exact}) derived in the no-passing case (the $R\to\infty$ limit). Still,
this term reflects the fact that collisions occur only with slower clusters
and that the collision rate is proportional to the velocity difference.  The
passing term is exact since the concentration of slowed down cars with
intrinsic velocity $v$ equals $P_0(v)-P(v,t)$.

In contrast with the no-passing case, the process is now stochastic in nature
and the system approaches a nontrivial steady state. Setting the time
derivative in Eq.~(\ref{pvt-p}) to zero we see that the steady state cluster
velocity distribution $P(v)\equiv P(v,t=\infty)$ satisfies the integral
equation
\begin{equation}
\label{steady}
P(v)\left[1+R\int_0^v dv'(v-v')P(v')\right]=P_0(v). 
\end{equation}
Given the intrinsic velocity distribution this relation gives the final
cluster velocity distribution only implicitly. In contrast, the inverse
problem is simpler as knowledge of the final distribution, the observed
quantity in real traffic flows, gives explicitly the intrinsic distribution.
We confirm that in the limit $R\to\infty$, all clusters move with the minimal
velocity, while in the limit $R\to 0$, all cars move with their intrinsic
velocity $P(v)\to P_0(v)$. 

The integral equation (\ref{steady}) can be transformed into a
differential one using the auxiliary function $Q(v)=R^{-1}+\int_0^v
dv'(v-v')P(v')$, from which \hbox{$P(v)=Q''(v)$}. Thus
Eq.~(\ref{steady}) becomes
\begin{equation}
\label{Qrho}
Q(v)Q''(v)=R^{-1}P_0(v).
\end{equation}
The boundary conditions are $Q(0)=R^{-1}$ and $Q'(0)=0$. The auxiliary
function $Q(v)$ gives a comprehensive description of the steady state.
Calculation of important quantities such as the flux $J$ requires knowledge
of $G(v)$, the car velocity distribution. This quantity satisfies
\hbox{$1=\int\!dv\, G(v)$} and \hbox{$J=\int\! dv \, v \, G(v)$}. Following
an involved calculation that requires solution of a higher order velocity
distribution \cite{bk1}, the car velocity distribution is derived explicitly
in terms of $Q(v)$
\begin{equation}
\label{g}
G(v)=P(v)\left[1+R\int_v^\infty dw\,P_0(w)\int_v^w {du\over [RQ(u)]^2}\right].
\end{equation}
Hence, for arbitrary intrinsic velocity distributions, the entire
steady state problem is reduced to the nonlinear second order
differential equation (\ref{Qrho}). Given $Q(v)$, steady state
distributions such as $P(v)$ and $G(v)$ can be calculated
using the explicit formulas above.

Except for a few special cases, one can not solve the differential
equation (\ref{Qrho}) analytically. Nevertheless, the formal solution 
above can be used to evaluate generic features of the flow. The
dimensionless collision number $R$ is extremely useful. For low
collision numbers, a perturbation solution in powers of $R$ can be
constructed, as the steady state differs weakly from the initial
state. For high collision numbers, a boundary layer analysis is
possible as sufficiently small velocities are not affected by
collisions. These two limits are quantitatively analyzed as follows. 

\subsection{Low Collision Numbers}

The flow characteristics in the collision-controlled regime, $R\ll 1$,
can be analyzed systematically as a perturbation series in $R$. For
example, the cluster velocity distribution and the car velocity distribution
read 
\begin{eqnarray}
\label{pv1}
&&P(v)\cong P_0(v)\left[1-R\int_0^v\!\!\!dv'(v-v')P_0(v')\right],\\
\label{gv1}
&&G(v)\cong P_0(v)\left[1+R\int_0^\infty dv' (v'-v)P_0(v')\right].\nonumber
\end{eqnarray}
Consequently, average quantities such as the flux and the average
cluster size vary linearly in $R$ in this free flow regime,
$J=J_0-{\rm const}\times R$, and $\langle m\rangle=1+{\rm const}\times
R$. The proportionality constant in the case of the flux equals the
variance in the initial velocity distribution, indicating that the
larger the initial velocity fluctuations, the larger the reduction in
the flux.

Therefore, weakly interacting ``laminar'' flows arise in the $R\to 0$
limit.  Technically, the steady state remains close to the initial
state and a perturbation series in the collision number is possible.
Here, the assumptions made in our theory are justified, as the cluster
sizes are small, and at the leading order, a simplified model were all
cars in the cluster can pass coincides with a more realistic model
where only the first few cars may pass. In fact, a basic prediction of
the model, namely linear growth of the average cluster size with the
flux is consistent with empirical data, obtained from observations of
traffic flows in a secondary rural road in Los Alamos, New Mexico
\cite{bk1}.

\subsection{High Collision Numbers}

The limit of high collision numbers corresponds to dense, congested flows
where large clusters form. The analysis in this passing-controlled regime is
more subtle since the condition \hbox{$R\int_0^v dv' (v-v')P_0(v')\ll 1$} is
satisfied only for small velocities.  No matter how large $R$ is,
sufficiently slow cars are not affected by collisions, and $P(v)$ is still
given by Eq.~(\ref{pv1}) when $v\ll v^*$. The threshold velocity $v^*\equiv
v^*(R)$ is estimated from \hbox{$R\int_0^{v^*} dv(v^*-v)P_0(v)\sim 1$}.

In the limit $R\to\infty$ limit, statistics of the slowest cars dominate the
flow.  Again, it is useful to consider intrinsic distributions with an
algebraic small velocity form (\ref{piv}). For such distributions, the
threshold velocity decreases with growing $R$ according to
\begin{equation}
\label{vstar}
v^* \sim R^{-{1\over   \mu+2}}.
\end{equation}
One can show that the flux is proportional to this velocity $J\sim v^*$.  For
$v\gg v^*$, the collision integral in Eq.~(\ref{steady}) dominates over the
constant factor and \hbox{$R P(v)\int_0^v dv' (v-v') P(v')\sim v^{\mu}$}.
Anticipating an algebraic behavior for the cluster velocity distribution,
$P(v)\sim R^{\sigma}v^{\delta}$ when $v\gg v^*$, gives different answers
dependent on whether the cutoff exponent $\mu$ is positive or negative.  The
leading behavior for $v\gg v^*$ can be summarized as follows
\begin{equation}
\label{pvlead}
P(v)\sim\cases{(v^*)^\mu(v/v^*)^{\mu-1}&$\mu<0$;\cr 
               (v/v^*)^{-1}[\ln(v/v^*)]^{-{1\over 2}}&$\mu=0$;\cr
               (v^*)^\mu (v/v^*)^{{\mu\over 2}-1}&$\mu>0$.}
\end{equation}
On the other hand, $P(v)\cong P_0(v)$ for $v\ll v^*$. This shows that the
velocity distribution develops a boundary layer structure, the size of which
vanishes in the infinite collision number limit. Inside the boundary layer,
the velocity distribution is only marginally lower than its initial values,
while the bulk of the velocities are strongly suppressed. Similar to the
threshold velocity $v^*$, macroscopic characteristics of the flow depend
algebraically on $R$. For example, the average cluster size is
\begin{equation}
\label{mav}
\langle m\rangle\sim\cases{R^{(\mu+1)/(\mu+2)} &$\mu<0$;\cr
                           (R/\ln R)^{1/2}     &$\mu=0$;\cr
                           R^{1/2}             &$\mu> 0$.}
\end{equation}
Two distinct regimes of behavior emerge.  For $\mu>0$, car-cluster collisions
dominate while for $\mu<0$ cluster-cluster collisions dominate.
Interestingly, in the cluster-cluster dominated regime, $\langle m\rangle\sim
R^{\alpha}$ with the scaling exponent \hbox{$\alpha=(\mu+1)/(\mu+2)$}
as in the no-passing case (\ref{cvscl}).  Thus in the passing case the
cutoff exponent $\mu$ also plays an important role in characterizing the
behavior.  Moreover, the steady state behavior is much richer than that found
for the clustering kinetics. 

Despite the simplifying assumptions, the model results in realistic behavior.
The overall picture is both familiar and intuitive: due to the presence of
slower cars, clusters form and the overall flux is reduced.  For heavy
traffic, the characteristics of the flow are solely determined by the
distribution of slow cars. A single dimensionless parameter, the collision
number $R$, ultimately determines the nature of the steady state.

\section{The Maxwell model}

While a comprehensive analysis of the steady state velocity
distributions is possible using the approximate kinetic theory
(\ref{pvt-p}), other important questions such as the relaxation
toward the steady state and the nature of the cluster size
distribution \cite{Gavrilov} remain unanswered. To address these
issues we consider a further approximation where the collision rate is
taken to be uniform \cite{bk3,ik1}. This approximation, known as the Maxwell
model, is very useful in kinetic theory \cite{ernst} and it has been
recently applied to granular gases as well 
\cite{gran1,gran2,gran3}. In our case, it allows for a complete exact
solution of the time dependent behavior, and additionally, it leads
to closed evolution equations for the cluster-size distribution.

\subsection{Relaxation}

In the Maxwell approximation, the collision rate $v-v'$ in the Boltzmann
equation (\ref{pvt-p}) is replaced by a constant factor which we set equal
to unity. The corresponding rate equation for the cluster velocity
distribution reads
\begin{eqnarray}
\label{pvt-m}
{\partial P(v,t)\over \partial t}&=&R^{-1}\left[P_0(v)-P(v,t)\right]\\
&-&P(v,t)\int_0^v dv' P(v',t).\nonumber
\end{eqnarray}
Again, the analysis is performed via a properly defined auxiliary function,
$Q(v,t)=\int_0^v dv' P(v',t)$. The constant collision rate results in simpler
differential equations, that are only first order in the velocity. The analog
of Eq.~(\ref{Qrho}) is the integrable steady state equation
$Q(v)Q'(v)=R^{-1}P_0(v)$.  The resulting steady state properties are governed
by $R$, with a boundary layer structure in the large $R$ regime. The
quantitative characteristics are somewhat different and for example the
threshold velocity decays with $R$ according to $v^*\sim R^{-1/(\mu+1)}$
rather than Eq.~(\ref{vstar}).  If, however, the collision rate is properly
chosen, namely set equal to $\langle v\rangle$ rather than unity, we recover 
Eq.~(\ref{vstar}).

Furthermore, the complete time dependence can be obtained analytically
by integrating the partial differential equation $Q_t=R^{-1}Q_v-QQ_v$.
In general, the relaxation is exponential $P(v,t)-P(v,t=\infty) \sim
f(v) e^{-t/\tau(v)}$, with $\tau(v)=R\left[1+2RI_0(v)\right]^{-1/2}$
where $I_0(v)=\int_0^v dv' P_0(v')$.  The relaxation time depends on 
the velocity and the collision number according to
\begin{equation}
\label{tauv}
\tau(v)\sim\cases{R                &$v\ll v^*$;\cr 
                  [R/I_0(v)]^{1/2} &$v\gg v^*$.}
\end{equation}
While small velocities are governed by practically fixed relaxation times,
large velocities are characterized by velocity dependent decay rates.
Furthermore, a large range of relaxation scales exists, $R^{1/2}<\tau< R$,
with larger scales corresponding to smaller velocities. Further analysis
shows that the same relaxation times underlie the car velocity distribution.
We expect that while the predictions of the Maxwell model are only
approximate, it correctly predicts the existence of a spectrum of relaxation
time scales, and that the qualitative nature of the time dependent behavior
generally holds.

\subsection{The size distribution}

The size distribution obeys closed evolution equations in the Maxwell model
and can be solved exactly \cite{bk3,ik1}. It can also be used to address the
nature of the passing mechanism.  To demonstrate this we consider the model
where only the next-to-leading car in the cluster may pass and resume driving
with its intrinsic velocity.  From numerical simulations of this model, we
find two distinct phases.  In the laminar regime, clusters are generally
small, specifically the cluster size distribution is exponentially suppressed
for sufficiently large sizes. When the collision number exceeds a certain
threshold, an infinite cluster is formed, i.e., a finite fraction of the cars
in the system are in the cluster behind the slowest car.  Furthermore, in
this jammed phase the size distribution of finite clusters has a fat tail
close to a power-law, $P_m\sim m^{-\tau}$, with $\tau\approx 2$.

In the Maxwell model framework, the cluster size distribution $P_m(t)$
obeys a closed system of rate equations
\begin{eqnarray} 
\label{simple} 
&&{d P_m\over dt}
=R^{-1} [P_{m+1}-P_m]-c\,P_m+{1\over 2}\sum_{i+j=m} P_iP_j,\\
\label{simple1} 
&&{d P_1\over dt}=R^{-1} [P_2-P_1+c]-c\,P_1.
\end{eqnarray}
These equations were derived by enumerating all possible ways in which
clusters evolve. For instance, consider Eq.~(\ref{simple1}).
Collisions reduce the density of single cars, and the collision rate
is clearly equal to $c$, as it is velocity-independent.  The escape
term in Eq.~(\ref{simple1}) is understood by observing that the rate
of return of single cars into the system is $2P_2+\sum_{j\geq 3}
P_j=P_2-P_1+c$.  Here $P_2$ is singled out since passing transforms it
into two single cars while an escape from larger clusters produces
only one freely moving car.

Similar equations were previously studied in the context of
aggregation-fragmentation processes \cite{ps,barma,satya}.  Utilizing the
approach of Ref.~\cite{ps} we find that a phase transition occurs at $R_c=1$
\cite{ik1}.  For large $m$, the steady state size distribution is
\begin{equation}
\label{Pmlarge}
P_m\sim \cases{m^{-3/2}\,\left[1-(1-R)^2\right]^m  &$R<1$,\cr
               m^{-5/2}                            &$R\geq 1$.} 
\end{equation}
Hence in the laminar regime, the size distribution decays exponentially in
the large size limit.  In the congested phase, the size distribution has a
power law tail, and in addition there is an infinite cluster that contains
the following finite fraction of cars in the system:
\begin{equation}
\label{I}
I=\cases{0,        &$R<1$;\cr
         1-R^{-1}, &$R>1$.}
\end{equation}
Interestingly, this phase transition is similar to phase transitions
in driven diffusive systems {\em without} passing
\cite{Benjamini,krug,Evans,Ktitarev,Krug}.  Furthermore, The formation
of an infinite cluster is reminiscent of Bose-Einstein condensation
\cite{Evans,barma}.

\section{Discussion}

The most important question raised by the above results concerns the validity
of the ``mean-field'' Boltzmann equation (\ref{pvt-p}). Although passing is a
stochastic mixing mechanism that diminishes correlations between the
velocities and the positions of the cars, such correlations do exist, and it
will be interesting to determine whether quantitative predictions such as the
scaling behaviors (\ref{vstar}) and (\ref{mav}) are altered by spatial
correlations. Similarly, the collision term in Eq.(\ref{simple}) is written
in a mean-field spirit and that may be the reason for the discrepancy between
the theoretical prediction $\tau=5/2$ and the numerically observed value
$\tau\approx 2$ of the decay exponent $P_m\sim m^{-\tau}$.

The primary feature of our model is {\em quenched disorder}, which
manifests itself in the random assignment of intrinsic
velocities. Road conditions (construction zones, intermittent passing
zones, turns, hills, etc.)  present another source of quenched
randomness in actual roads \cite{other}, which is ignored
in our model.  Quenched disorder significantly affects characteristics
of many-particle systems, especially in low spatial dimensions
\cite{bg}.  We have seen that this general conclusion clearly applies to
our one-dimensional traffic model.  Little is known analytically on 
the influence of the spatial disorder.

Finally, one may modify the passing rule so that when a car overtakes
a slow car, it acquires a new velocity drawn from the distribution
$P_0(v)$ rather than a pre-assigned velocity \cite{ik2}. This
elementary zero-memory model remains highly non-trivial even in the
collision-controlled limit $R\to 0$ where  clustering can be
disregarded. The fate of the system is again determined by the
behavior of the intrinsic velocity distribution near its lower cutoff. 
If $P_0(v)$ vanishes in this limit, the system reaches a steady state, 
otherwise, the system evolves indefinitely.  Specifically, for
intrinsic distributions with an algebraic small velocity tail
(\ref{piv}) the long-time asymptotics of the average velocity reads
\begin{equation}
\label{av}
\langle v(t)\rangle \sim\cases{{\rm const}  &  $\mu>0$;\cr
               (\ln t)^{-1} &  $\mu=0$;\cr
               t^\mu        &  $-1<\mu<0$.\cr}
\end{equation}
These results were derived in a simplified Boltzmann framework.  In
particular, the most interesting behavior in the evolving regime was obtained
by assuming that as $t\to\infty$, cars can be divided into two groups, the
small group of ``active'' cars which move with velocities $v\sim 1$ and the
vast majority of ``creeping'' cars that hardly move at all.  We then ignored
collisions between creeping cars (since their relative velocity is very
small) and collisions between active cars (since their density is small).
Thence, the velocity distribution of active cars obeys a linear
Boltzmann-Lorentz equation which was solved to give (\ref{av}).  Comparison
with results of molecular dynamics simulations suggests that the mean-field
theory description is asymptotically exact.  It will be interesting to
confirm this result rigorously.

\vspace{.2in}
\noindent 
We are thankful to our collaborators Slava Ispolatov and Sid Redner. 
This research was supported by DOE (W-7405-ENG-36) and
NSF(DMR9978902).

\end{multicols}
\end{document}